\documentclass[letter,twocolumn]{jpsj3}
\usepackage{txfonts}
\usepackage{amsmath}
\usepackage{bm}
\usepackage{braket}
\usepackage{amssymb}
\usepackage{color}
\usepackage{ulem}

\newcounter{num}
\newcommand{\rnum}[1]{\setcounter{num}{#1} \roman{num}}
\allowdisplaybreaks[4]

\title{
  Nonlinear Transverse Magnetic Susceptibility\\ under Electric Toroidal Dipole Ordering
}

\author{
  Akane Inda 
  and 
  Satoru Hayami 
  }
\inst{
  Department of Physics, Hokkaido University, Sapporo 060-0810, Japan\\ 
} 

\abst{
An electric toroidal dipole (ETD) moment is one of the fundamental dipole moments as well as electric and magnetic ones.
Although it directly couples to neither an electric nor magnetic field due to its spatial inversion and time-reversal parities, its ordered state leads to unconventional transverse responses of the conjugate physical quantities. 
We here theoretically investigate nonlinear transverse magnetic susceptibility under the ETD ordering.
By performing a self-consistent mean-field calculation for a five $d$-orbital model under a tetragonal crystalline electric field and using the nonlinear Kubo formula, we show that a third-order transverse magnetic susceptibility corresponding to a uniform magnetization perpendicular to the external magnetic field becomes nonzero once the ETD moment is ordered under tetragonal crystalline electric field.
Moreover, we find that spin-orbital entanglement and a low-lying first excited crystal-field level are important for realizing large transverse responses. 
}
\begin{document}
\maketitle
Electronic ordered phases, which arise from spontaneous symmetry breaking due to electron correlations, show a variety of functionalities in materials~\cite{Possible_Species_of_Ferromagnetic_Ferroelectric_and_Ferroelastic_Crystals, Broken_symmetries_non-reciprocity_and_multiferroicity}. 
The appearance of such functionalities is qualitatively understood from its transformation in terms of spatial inversion ($\mathcal{P}$) and time-reversal ($\mathcal{T}$) symmetries.  
For example, ferroelectricity appears when the $\mathcal{P}$ symmetry is broken~\cite{Possible_Species_of_Ferroelectrics}, while ferromagnetism appears when the $\mathcal{T}$ symmetry is broken. 
In addition, ferrotoroidicity related to multiferroic responses appears when both $\mathcal{P}$ and $\mathcal{T}$ symmetries are broken~\cite{Some_symmetry_aspects_of_ferroics_and_single_phase_multiferroics, Advances_in_magnetoelectric_multiferroics}. 
Microscopically, they are characterized by a ferroic alignment of dipole moments; ferroelectricity, ferromagnetism, and ferrotoroidicity are described by a ferroic alignment of electric, magnetic, and magnetic toroidal dipoles among the electronic degrees of freedom, respectively, based on the multipole description~\cite{Spaldin_0953-8984-20-43-434203, Hlinka_PhysRevLett.113.165502, hayami2018microscopic, Classification_of_atomic-scale_multipoles, Watanabe_PhysRevB.98.245129, Watanabe_PhysRevB.98.220412, classification_in_122_magnetic_point_groups}. 

\begin{figure}[t]
\label{f1}
\centering
\includegraphics[width=\linewidth]{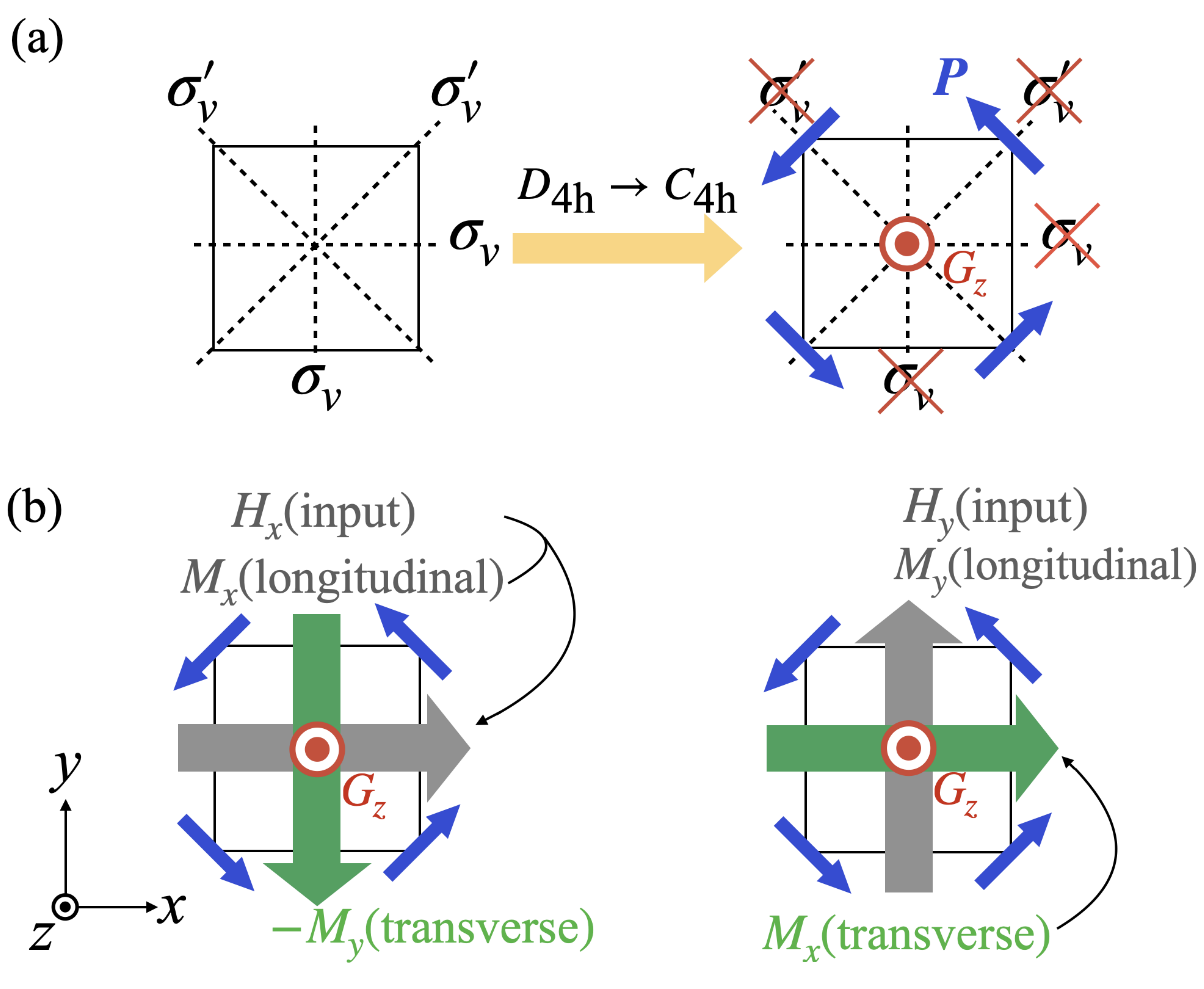}
\caption{(Color online) 
(a) Schematic structures under the point groups $D_{\text{4h}}$ and $C_{\text{4h}}$. 
When the symmetry is lowered from 
$D_{\text{4h}}$ (left panel) to $C_{\text{4h}}$ (right panel) owing to the breakings of vertical mirror symmetries $\sigma_v$ and $\sigma'_v$ in the presence of the vortex-type electric polarization $\bm{P}$ (blue arrows), the ETD moment $G_z$ is induced along the $z$ direction. 
(b) Schematic pictures of the transverse responses against an external magnetic field under the ETD ordering. 
The transverse magnetization $-M_y$ ($M_x$) as well as the longitudinal one $M_x$ ($M_y$) is induced by the magnetic field $H_x$ ($H_y$).
\label{fig: fig1}
}
\end{figure}

Another dipole moment, which is referred to as an electric axial moment, exists neither $\mathcal{P}$ nor $\mathcal{T}$ symmetry breaking~\cite{Symmetry_Guide_to_Ferroaxial_Transitions}.
The axial moment microscopically corresponds to a ferroic alignment of an electric toroidal dipole (ETD) moment.
This ordered state appears when mirror symmetry parallel to the ETD moment is broken in the system.
We show an example of the tetragonal-lattice case in Fig.~\ref{fig: fig1}(a); when the mirror symmetries of the square object ($\sigma_v$ and $\sigma'_v$) in the left panel of Fig.~\ref{fig: fig1}(a) are broken according to the emergence of vortex-type electric polarization $\bm{P}$ denoted by the blue arrows in the right panel of Fig.~\ref{fig: fig1}(a), 
the symmetry of the system reduces from $D_{\rm 4h}$ to $C_{\rm 4h}$, and then, the $z$ component of the ETD, $G_z$, is induced. 
In such a situation, intriguing transverse responses of the conjugate physical quantities are expected owing to mirror symmetry breaking~\cite{cheong2021permutable}, such as the 
spin current generation~\cite{Roy_PhysRevMaterials.6.045004, Hayami_Ferro-axial_spincurrent} and the antisymmetric thermopolarization~\cite{thermopolarization}. 
Uncovering such ETD-related physical phenomena is useful to further explore functionalities characteristic of the ETD ordering in materials, such as CaMn$_7$O$_{12}$~\cite{ETD_CaMn7O12}, RbFe(MoO$_4$)$_2$~\cite{ETD_RbFeMoO42,Hayashida_PhysRevMaterials.5.124409}, and NiTiO$_3$.~\cite{Visualization_of_ferroaxial_domains_in_an_order-disorder_type_ferroaxial_crystal, Hayashida_PhysRevMaterials.5.124409, yokota2022three}

In the present study, we theoretically examine another transverse response under the ETD ordering by focusing on the nonlinear magnetic susceptibility (MS). 
To extract important ingredients for nonlinear transverse MS (TMS), we construct a minimal $d^1$ model on a square lattice, which includes the atomic-scale ETD degree of freedom consisting of the outer product of spin and orbital angular momenta. 
Next, we perform the mean-field calculations for the model Hamiltonian incorporating the effects of the spin--orbit coupling (SOC) and the crystalline electric field (CEF).
Then, we discuss the behavior of the third-order TMS under the atomic-scale ETD ordering based on the nonlinear Kubo formula~\cite{Kubo_formula}. 
By analyzing the expression of the third-order TMS, we find that coupling between spin and orbital angular momenta under the tetragonal CEF and the small energy difference between the ground-state and first excited-state CEF levels are key ingredients to enhance the responses. 
We also discuss the case when the order parameter is characterized by an electric hexadecapole without the spin operator, which is another degree of freedom corresponding to the ferroaxial moment in the tetragonal model.

We consider five $d$ orbitals, ($\phi_u$, $\phi_v$, $\phi_{yz}$, $\phi_{zx}$, $\phi_{xy}$) for $u=3z^2-r^2$ and $v=x^2-y^2$, with $d^1$ configuration under the point group $D_{\text{4h}}$. 
Each atomic site is separated by the lattice constant $a=1$ on the two-dimensional plane to form a simple square lattice.
The model Hamiltonian under $D_{4\rm h}$ is given by
\begin{align}
\label{e_Hamiltonian}
\hat{\mathcal{H}}^{\rm loc} 
&= \hat{\mathcal{H}}_{\text{CEF}}
+
\hat{\mathcal{H}}_{\text{SOC}}.
\end{align}
The first term $\hat{\mathcal{H}}_{\text{CEF}}$ represents the CEF Hamiltonian, which split the atomic-energy level under the tetragonal CEF. 
We set the CEF parameters to satisfy $
E_{xy}-E_{yz}=1.6$, $
E_u-E_{yz}=1.9$, 
$E_v-E_{yz}=2.8$, and $E_{zx}=E_{yz}=0$, where $E_{\alpha}$ is the atomic-energy level for $\alpha=(u, v, yz, zx, xy)$. 
The second term $\hat{\mathcal{H}}_{\text{SOC}}$ represents the atomic SOC, which is represented by 
\begin{align}
\label{e_SOC}
\hat{\mathcal{H}}_{\text{SOC}}
&=\lambda \sum_{m} \hat{\bm{l}}_{m} \cdot \hat{\bm{s}}_{m},
\end{align} 
where $\hat{\bm{l}}_{m}$ and $\hat{\bm{s}}_{m}$ are the orbital and spin angular momentum operators at site $m$, respectively.

In the five $d$-orbital Hilbert space, there are independent 100 electronic degrees of freedom: 25 degrees of freedom in spinless space and 75 degrees of freedom in spinful space.
Among them, only the four degrees of freedom belonging to the ${\rm A}^+_{2g}$ representation are compatible with the electric ferroaxial moment along the $z$ direction. 
Based on the augmented multipole description~\cite{kusunose2020complete, kusunose2022generalization}, such four degrees of freedom are described by the electric hexadecapole $Q^{\alpha}_{4z} \propto xy (x^2-y^2)$ in spinless space and the ETD $G_z$, the electric toroidal octopole $G^\alpha_{z}$, and another electric hexadecapole $Q'^{\alpha}_{4z}$ in spinful space~\cite{Hayami_Ferro-axial_spincurrent}. 
In particular, we focus on the $G_z$ ordering in the following analyses, whose operator is defined in the atomic scale by~\cite{Hayami_Ferro-axial_spincurrent, Ferroaxial_moment_induced_by_vortex_spin_texture, hoshino2022spin} 
\begin{align}
\label{eq: Gz}
\hat{G}_{zm} =  \big(\hat{\bm{l}}_m \times \hat{\bm{s}}_m\big)^z. 
\end{align}
It is noted that $\langle \hat{G}_{zm} \rangle$ becomes nonzero without the magnetic moments, i.e., $\langle \hat{\bm{l}}_{m} \rangle=\langle \hat{\bm{s}}_{m} \rangle=\bm{0}$, where $\braket{\cdots}$ represents the statistical average in $d^1$ configuration.

In order to discuss the situation where the primary order parameter corresponds to $\hat{G}_{zm}$, we phenomenologically introduce an effective interaction between the ETD at sites $m$ and $n$, which is represented by
\begin{align}
\label{e_interaction}
\hat{\mathcal{H}}_{\text{int}}
&= -\sum_{m, n}J_{mn}  \hat{G}_{zm} \hat{G}_{zn},
\end{align} 
where $J_{mn}$ is the coupling constant for the nearest-neighbor sites. We apply the mean-field approximation for this term as
\begin{align}
\label{e_mf}
\hat{\mathcal{H}}_{\text{int}}^{\text{MF}}
=
-JG_z\sum_{m}\hat{G}_{zm}+ \  {\rm (const.)},
\end{align}
where we set the ferroic interaction $J \equiv 4J_{mn}>0$ and $G_z \equiv \Braket{\hat{G}_{zm}}$ by supposing the single-sublattice structure; we omit the site index hereafter. 
$J$ is the energy unit of the model.

\begin{figure}[t]
\centering
\includegraphics[width=\linewidth]{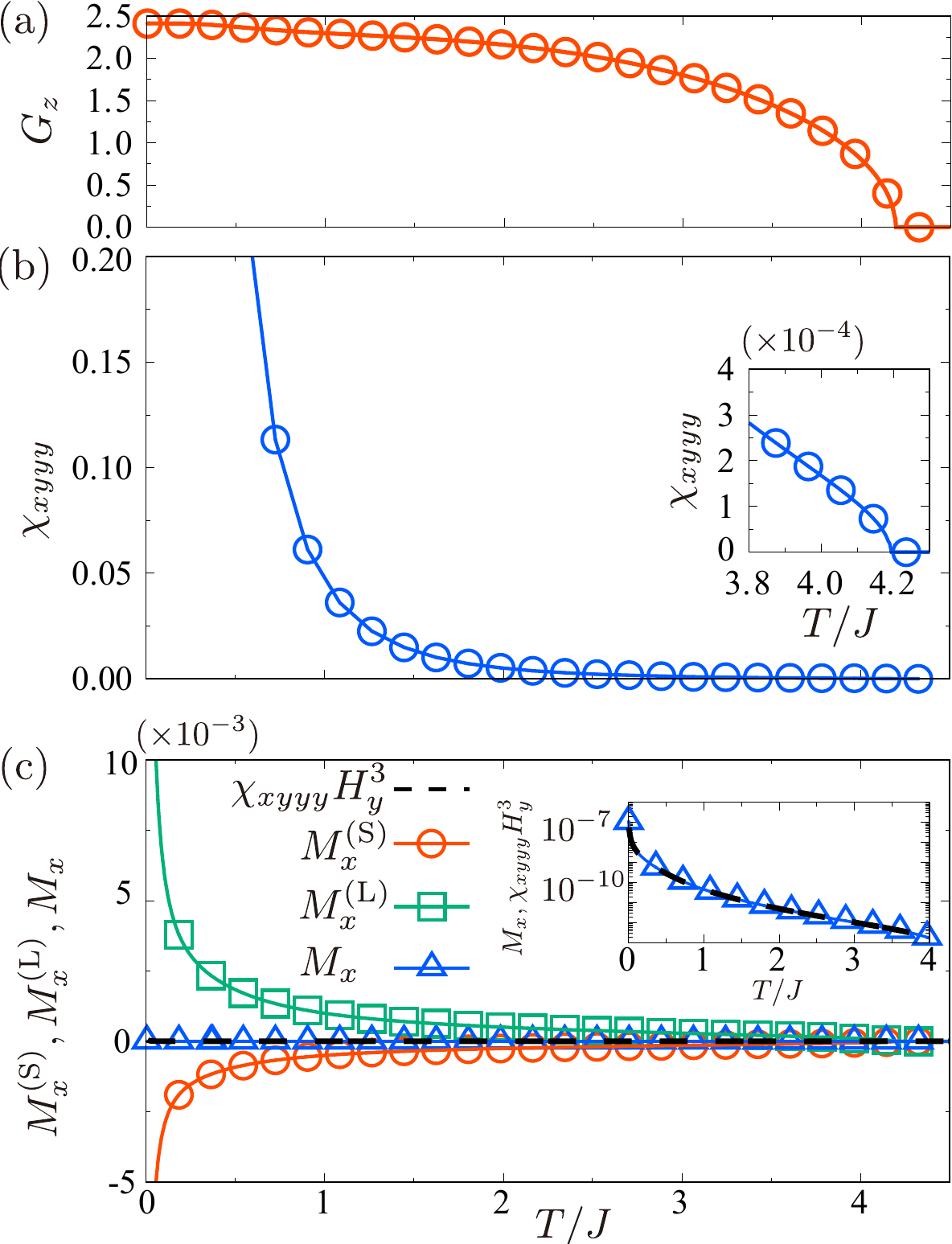}
\caption{(Color online) 
Temperature dependence of (a) $G_z$ and (b) $\chi_{xyyy}$ for $\hat{\mathcal{H}}$ at $\lambda=0.2$~\cite{comment_data}.
The inset of (b) represents the behavior near the transition temperature. 
(c) Temperature dependence of $
M^{\rm (S)}_x$, $M^{\rm (L)}_x$, and $ 
M_x$ for $\hat{\mathcal{H}}+ \hat{\mathcal{H}}_{\text{Zeeman}}$ at $\lambda=0.2$ and $H_y=10^{-3}$.
For reference, the data in (b) is also plotted by the dashed curve in (c). 
The inset of (c) represents the enlarged figure to show the behaviors of $
M_x$ and $\chi_{xyyy}H_y^3$.
}
\label{f2}
\end{figure}

By performing the self-consistent mean-field calculations for the total Hamiltonian $\hat{\mathcal{H}}=\hat{\mathcal{H}}^{\rm loc} +\hat{\mathcal{H}}_{\text{int}}^{\text{MF}}$, the transition to the ETD ordering occurs at a finite temperature; the point group symmetry reduces from $D_{\rm 4h}$ to $C_{\rm 4h}$.
Figure~\ref{f2}(a) shows the temperature ($T$) dependence of $G_z$ at $\lambda=0.2$~\cite{comment_data}. 
As shown in Fig.~\ref{f2}(a), the transition from the paramagnetic state to the ETD-ordered state occurs at $T\simeq 4.2$. 
Note that under the ETD ordering with $G_z \neq 0$, $\langle \hat{G}^{\alpha}_z \rangle $ and $\langle \hat{Q}'^{\alpha}_{4z} \rangle$ are also induced but $\langle \hat{Q}^{\alpha}_{4z} \rangle=0$.

Under such an ETD ordering, we investigate the transverse response against the external magnetic field, whose Hamiltonian is given by
\begin{align}
\label{e_Zeeman}
\hat{\mathcal{H}}_{\text{Zeeman}}= -\mu_\text{B} \bm{H}\cdot 
\hat{\bm{\mu}},
\end{align} 
where we consider the effect of the Zeeman coupling for simplicity. 
$\hat{\bm{\mu}} 
=\hat{\bm{l}} 
+ 2 \hat{\bm{s}} 
$ and $\bm{H}=(H_x, H_y, H_z)$.
We set the Bohr magneton as unity, i.e., $\mu_\textrm{B}=1$. 
The MS is given as the expansion coefficient when the magnetization $\bm{M}$ is expanded in terms of $\bm{H}$ as follows:
\begin{align}
\label{e40}
M_\eta 
= 
\chi_{\eta\mu}H_\mu+\chi_{\eta\mu\nu \kappa}H_\mu H_\nu H_\kappa+\cdots,
\end{align} 
for $\eta, \mu, \nu, \kappa = x,y,z$. 
$\chi_{\eta\mu}$ and $\chi_{\eta\mu\nu \kappa}$ stand for the linear and third-order MS, respectively; the even-order contribution vanishes owing to the $\mathcal{T}$ symmetry.
When $G_z \neq 0$ under the point group $C_{\text{4h}}$, the following tensor components can additionally become nonzero compared to the case under $D_{\rm 4h}$ from the symmetry viewpoint~\cite{classification_in_122_magnetic_point_groups}: $\chi_{xy}=-\chi_{yx}$ and $\chi_{xyyy}=-\chi_{yxxx}$, $\chi_{xxxy}=-\chi_{yxyy}$, and $\chi_{xyzz}=-\chi_{yzzx}$.
These tensor components correspond to the transverse responses. 
For example, $\chi_{xy}$ ($=-\chi_{yx}$) [$\chi_{xyyy}$ ($=-\chi_{yxxx}$)] means that the magnetization is induced perpendicular to the external magnetic field; $-M_y$ ($M_x$) is induced by $H_x$ ($H_y$), as schematically shown in Fig.~\ref{fig: fig1}(b). 
It is, however, noted that the linear antisymmetric component of $\chi_{xy}$ ($=-\chi_{yx}$) vanishes in the linear response theory based on the Kubo formula:
$\chi_{xy}=-\chi_{yx}=0$. 
Meanwhile, the third-order transverse response remains, e.g., $\chi_{xyyy}=-\chi_{yxxx} \neq 0$. 
To demonstrate that, we especially focus on $\chi_{xyyy}$ and numerically evaluate it based on the nonlinear Kubo formula~\cite{Kubo_formula}. 

By supposing the static limit ($ \omega\rightarrow 0$ 
and then $\bm{q} \rightarrow \bm{0}$), we obtain the following expression given by
\begin{align}
\label{e45}
& \chi_{xyyy} 
=
\chi^{(\mathrm{\rnum{1}})}
+
\chi^{(\mathrm{\rnum{2}})}
+
\chi^{(\mathrm{\rnum{3}})}
+
\chi^{(\mathrm{\rnum{4}})}
+
\chi^{(\mathrm{\rnum{5}})},
\end{align}
where
\begin{align}
\label{e_sus_full}
\chi^{(\mathrm{\rnum{1}})} 
&= \frac{1}{6} \sum_{ijkl}^{\xi_i=\xi_j=\xi_k=\xi_l}\rho'''_i\mathcal{M}_{ijkl}^{xyyy},\\
\chi^{(\mathrm{\rnum{2}})} 
&=
-\Bigg\{
\sum_{ijkl}^{\xi_i=\xi_k=\xi_l\neq\xi_j}
\frac{1}{\xi_{ij}}\Bigg(
\frac{\rho''_i}{2}
+\frac{\rho'_i}{\xi_{ij}}
+\frac{\rho_{ij}}{\xi_{ij}^2}
\Bigg)\nonumber\\
&\qquad+(i\leftrightarrow j)
+(j\leftrightarrow k)
+(j\leftrightarrow l)
\Bigg\}
\mathcal{M}^{xyyy}_{ijkl},\\
\label{eq: chi3}
\chi^{(\mathrm{\rnum{3}})}&=
\sum_{ijkl}^{\xi_i=\xi_j\neq\xi_k=\xi_l}\frac{1}{\xi_{ik}^2}
\Bigg\{
(\rho'_i+\rho'_k)
+2\frac{\rho_{ik}}{\xi_{ik}}
\Bigg\}\mathcal{M}_{ijkl}^{xyyy}  \nonumber \\
&+2\sum_{ijkl}^{\xi_i=\xi_k\neq \xi_j=\xi_l}\frac{1}{\xi_{ij}^2}
\Bigg(
\rho'_i+\frac{\rho_{ij}}{\xi_{ij}}\Bigg)
\Big(\mathcal{M}_{ijkl}^{xyyy}+\mathcal{M}_{ijlk}^{xyyy}\Big)
,\\
\chi^{(\mathrm{\rnum{4}})} 
&=\sum_{ijkl}^{\xi_i=\xi_j\neq\xi_k\neq\xi_l}
\frac{1}{\xi_{ik}\xi_{il}} \Bigg(
\rho'_i
+2\frac{\rho_{ik}} {\xi_{ik}}
-\frac{\rho_{lk}}{\xi_{lk}}\Bigg)\mathcal{M}_{ijkl}^{xyyy} \nonumber \\
&+2\sum_{ijkl}^{\xi_i=\xi_k\neq\xi_j\neq\xi_l} 
\frac{1}{\xi_{ij}\xi_{il}}
\Bigg(
\rho'_i
+
\frac{\rho_{il}}{\xi_{il}}
+
W_{ijl}
\Bigg)\Big(\mathcal{M}_{ijkl}^{xyyy}+\mathcal{M}_{ijlk}^{xyyy}\Big) \nonumber \\
& + \sum_{ijkl}^{\xi_k=\xi_l\neq\xi_i\neq\xi_j}
\frac{1}{\xi_{ik}}\Bigg(
\frac{\rho'_k}{\xi_{jk}}
-2\frac{\rho_{ik}}{\xi_{ij}\xi_{ik}}
\Bigg)\mathcal{M}_{ijkl}^{xyyy},
\\
\label{eq:chiV}
\chi^{(\mathrm{\rnum{5}})} &= 2 \sum_{ijkl}^{\xi_i\neq\xi_j\neq\xi_k\neq\xi_l}\Bigg\{\frac{1}{\xi_{ij}\xi_{ik}}
\Bigg(
\frac{\rho_{lk}}{\xi_{lk}}-\frac{\rho_{il}}{\xi_{il}}
\Bigg)
\mathcal{M}_{ijkl}^{xyyy}\Bigg\},
\end{align}
with $\mathcal{M}_{ijkl}^{xyyy}=\mu^x_{ij}\mu^y_{jk}\mu^y_{kl}\mu^y_{li}$. 
$\xi_j$ and $\rho_j$
are the $j$th eigenenergies and density matrices for $\hat{\mathcal{H}}$, respectively, and $\beta=1/T$ is the inverse temperature (The Boltzmann constant is set to be unity.); 
$\rho_j =e^{-\beta \xi_j}/Z$
($Z=\sum_i e^{-\beta \xi_i}$ is the partition function),
$\xi_{ij}= \xi_i-\xi_j $,
$\rho_{ij}=\rho_i-\rho_j$, $\rho'_i = \beta \rho_i$, $\rho''_i = \beta^2 \rho_i$, $\rho'''_i = \beta^3 \rho_i$, and $W_{ijl}=(\xi_l\rho_{ji}+\xi_j\rho_{il}+\xi_i\rho_{lj})/\xi_{ij}\xi_{jl}$.
We sort the energy levels to satisfy $\xi_1 \leq \xi_{2} \leq  \cdots  \leq \xi_{10}$.
Among the contributions, $\chi^{(\mathrm{\rnum{1}})}$ [$\chi^{(\mathrm{\rnum{5}})}$] includes only Curie-type (the transition between the state with the same energy) [van-Vleck-type (the transition between the states with different energies)] processes, while $\chi^{(\mathrm{\rnum{2}})}$, $\chi^{(\mathrm{\rnum{3}})}$, and $\chi^{(\mathrm{\rnum{4}})}$ represent other processes consisting of both Curie-type and van-Vleck-type processes; the superscript of the summation stands for the conditions for the eigenenergies.

Figure~\ref{f2}(b) shows the $T$ dependence of $\chi_{xyyy}$ at $\lambda=0.2$~\cite{comment_data}.
One finds that $\chi_{xyyy}$ becomes nonzero below the critical temperature [see also the inset of Fig.~\ref{f2}(b)].
When $T$ decreases, $\chi_{xyyy}$ increases due to the terms proportional to $\beta$, $\beta^2$, and $\beta^3$. 
This divergent behavior at $T \to 0$ is common to that of the longitudinal MS such as $\chi_{xx}$. 
Moreover, the relation of $\chi_{xyyy}=-\chi_{yxxx}$ is confirmed. 
In this way, the ETD ordering drives the third-order transverse magnetization against the external magnetic field.

\begin{figure}[t]
\centering
\includegraphics[width=\linewidth]{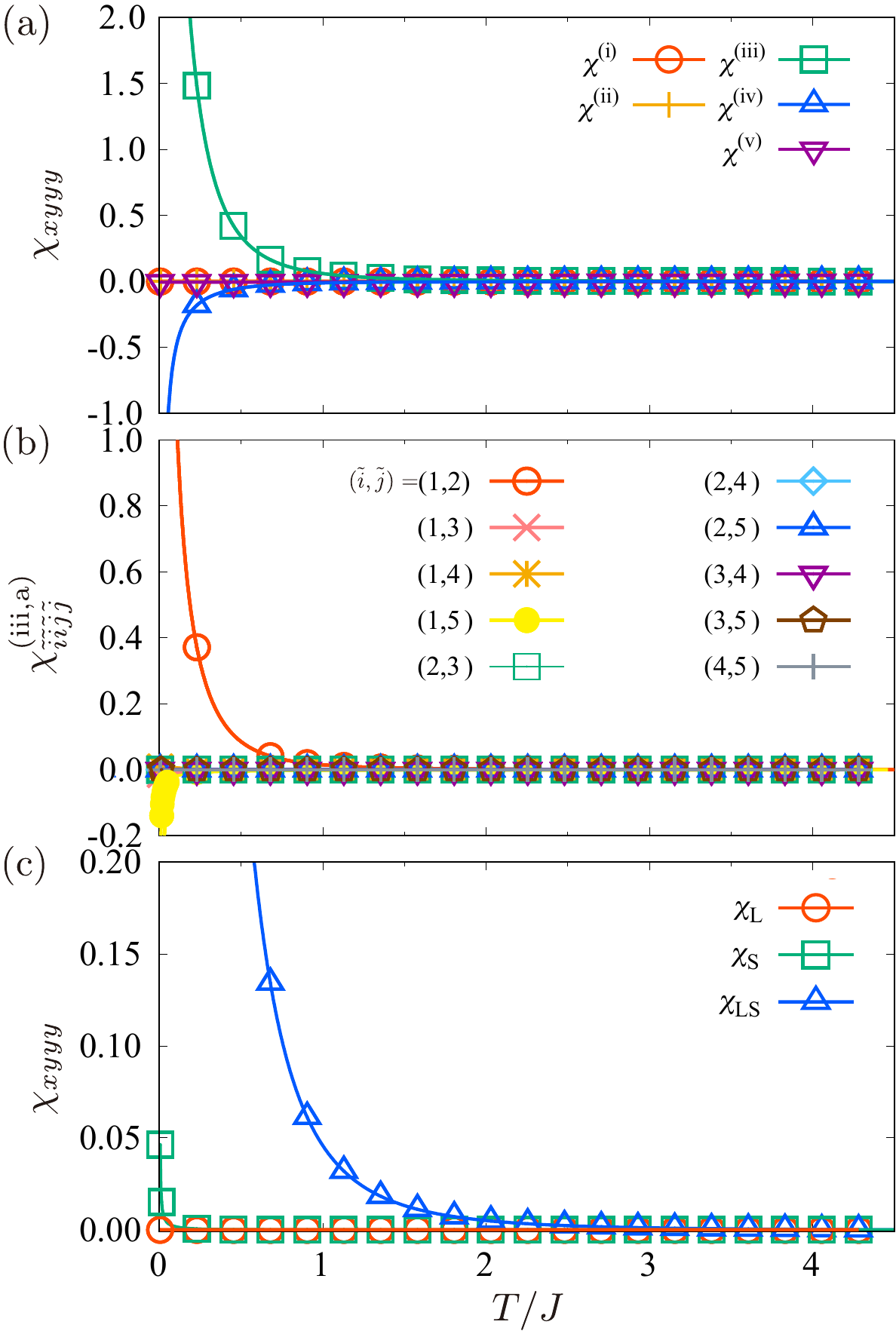}
\caption{(Color online) 
$T$ dependence of 
(a) ($\chi^{(\mathrm{\rnum{1}})}$, $\chi^{(\mathrm{\rnum{2}})}$, $\chi^{(\mathrm{\rnum{3}})}$, $\chi^{(\mathrm{\rnum{4}})}$, $\chi^{(\mathrm{\rnum{5}})}$) in Eqs.~(\ref{e_sus_full})--(\ref{eq:chiV}), 
(b) $\chi^{\mathrm{(iii,a)}}_{\tilde{i}\tilde{i}\tilde{j}\tilde{j}}$ in Eq.~(\ref{e_sus_3}), and 
(c) ($\chi_{\rm L}, \chi_{\rm S}, \chi_{\rm LS}$) in Eq.~(\ref{e_sus_ls}) at $\lambda=0.2$~\cite{comment_data}.
}
\label{f3}
\end{figure}

When $\chi_{xyyy}$ is obtained, the magnitude of the induced transverse magnetization is roughly estimated for small $H_y$ as $\chi_{xyyy} H^3_y$. 
Besides, one can directly evaluate the magnetization $M_x \equiv \langle \hat{\mu}_x \rangle $ by performing the mean-field calculations for $\hat{\mathcal{H}}+ \hat{\mathcal{H}}_{\text{Zeeman}}$ instead of $\hat{\mathcal{H}}$ by introducing small $H_y=10^{-3}$.
We compare them in Fig.~\ref{f2}(c), whose data indicate that $\chi_{xyyy}H_y^3$ agrees with $M_x$ [see also the inset of Fig.~\ref{f2}(c)]. 
$M_x$ takes the order of $10^{-7}$ at low temperatures, which is much smaller than the longitudinal magnetization $M_y$ to take around $10^{-1}$ for $H_y= 10^{-3}$ at $T =4.5\times 10^{-3}$. 
We find that this small transverse magnetization compared to the longitudinal one is attributed to the opposite sign of the orbital and spin components, which are represented by $M^{\rm (L)}_{x} \equiv \langle \hat{l}_x \rangle $ and $M^{\rm (S)}_{x} \equiv \langle \hat{s}_x \rangle$, respectively.
As shown in Fig.~\ref{f2}(c), the order of $M^{\rm (L)}_{x}$ and $M^{\rm (S)}_{x}$ is comparable with each other. 
Their opposite-sign tendency is qualitatively understood from the expression of $\hat{G}_z$ in Eq.~(\ref{eq: Gz}) with the coupling form of $\hat{l}_x \hat{s}_y-\hat{l}_y\hat{s}_x$, although it also depends on the CEF parameters; $+M^{\rm (L)}_{x}$ and $-M^{\rm (S)}_{x}$ are indirectly induced by nonzero $M^{\rm (S)}_{y}$ and $M^{\rm (L)}_{y}$ under $H_y$, respectively. 
In the present model, the minimum condition for the model parameters to induce net $M_x$ is given by $E_v \neq 0$ or  $E_{xy} \neq 0$.
Nevertheless, we show that the induced transverse magnetization can become larger when $M_x$ deviates from $\chi_{xyyy}H_y^3$ for large $H_y$, as will be discussed later. 

To further examine the behavior of $\chi_{xyyy}$ in Eq.~(\ref{e45}), we investigate its dominant transition processes.
By evaluating ($\chi^{(\mathrm{\rnum{1}})}$, $\chi^{(\mathrm{\rnum{2}})}$, $\chi^{(\mathrm{\rnum{3}})}$, $\chi^{(\mathrm{\rnum{4}})}$, $\chi^{(\mathrm{\rnum{5}})}$), one finds that $\chi^{(\mathrm{\rnum{3}})}$ gives the dominant contribution to $\chi_{xyyy}$, as shown in Fig.~\ref{f3}(a). 
It is noted that $\chi^{(\mathrm{\rnum{1}})}$ vanishes since it has only symmetric components. 
In addition, $\chi^{(\mathrm{\rnum{2}})}$ also vanishes, which is presumably accidental in the present model. 
We further decompose $\chi^{(\mathrm{\rnum{3}})}$ as 
\begin{align}
\label{e_sus_3}
\chi^{(\mathrm{\rnum{3}})}=
\sum_{\tilde{i}\tilde{j}}^{\xi_{\tilde{i}}\neq\xi_{\tilde{j}}}
\Big(\chi_{\tilde{i}\tilde{i}\tilde{j}\tilde{j}}^{(\mathrm{\rnum{3}}, a)}
+
\chi_{
\tilde{i}\tilde{j}\tilde{j}\tilde{i}}^{(\mathrm{\rnum{3}}, b)}
+
\chi_{\tilde{i}\tilde{j}\tilde{i}\tilde{j}}^{(\mathrm{\rnum{3}}, c)}\Big),
\end{align}
where the first term of Eq.~(\ref{e_sus_3}) corresponds to the first term in Eq.~(\ref{eq: chi3}) and the other two terms correspond to the second term in Eq.~(\ref{eq: chi3}). 
$\tilde{i}, \tilde{j}$ represent the non-degenerate energy-level indices; $\tilde{i}=(1, 2, 3, 4, 5)$ means the energy levels $i=\{(1,2), (3,4), (5,6), (7,8), (9,10)\}$, since there is twofold Kramers degeneracy in the energy levels.
By evaluating $(\chi_{\tilde{i}\tilde{i}\tilde{j}\tilde{j}}^{(\mathrm{\rnum{3}}, a)}, \chi_{\tilde{i}\tilde{j}\tilde{j}\tilde{i}}^{(\mathrm{\rnum{3}}, b)}, \chi_{\tilde{i}\tilde{j}\tilde{i}\tilde{j}}^{(\mathrm{\rnum{3}}, c)})$ for each $(\tilde{i},\tilde{j})$, we find that the dominant contribution arises from the process between the ground-state and first-excited-state levels in $\chi_{\tilde{i}\tilde{i}\tilde{j}\tilde{j}}^{(\mathrm{\rnum{3}}, a)}$, as shown in Fig.~\ref{f3}(b), where we only plot $\chi_{\tilde{i}\tilde{i}\tilde{j}\tilde{j}}^{(\mathrm{\rnum{3}}, a)}$ for simplicity.
In the end, the result indicates that there are two important observations to enhance $\chi_{xyyy}$: 
One is the energy-level structures to satisfy $\xi_i=\xi_j\neq\xi_k=\xi_l$ like $\chi^{(\mathrm{\rnum{3}})}$ in Eq.~(\ref{eq: chi3}) and the other is the small energy difference between the ground-state and first-excited-state levels. 

Next, we show that spin-orbital entanglement also plays an important role in enhancing $\chi_{xyyy}$. 
For this purpose, we perform a different decomposition of $\chi_{xyyy}$ from Eq.~(\ref{e_sus_3}) into three parts as
\begin{align}
\label{e_sus_ls}
\chi_{xyyy}= \chi_{\mathrm{L}}+\chi_{\mathrm{S}}+\chi_{\mathrm{LS}},
\end{align} 
where we replace $\mu \to l$ ($s$) in $\mathcal{M}^{xyyy}_{ijkl}$ for $\chi_{\mathrm{L}}$ ($\chi_{\mathrm{S}}$); $\chi_{\mathrm{LS}}$ represents the other components.
The data in Fig.~\ref{f3}(c) clearly indicate that $\chi_{\mathrm{LS}}$ is much larger than $\chi_{\mathrm{L}}$ and $\chi_{\mathrm{S}}$, which means that the coupling between spin and orbital components is important.

\begin{figure}[htbp]
\centering
\includegraphics[width=\linewidth]{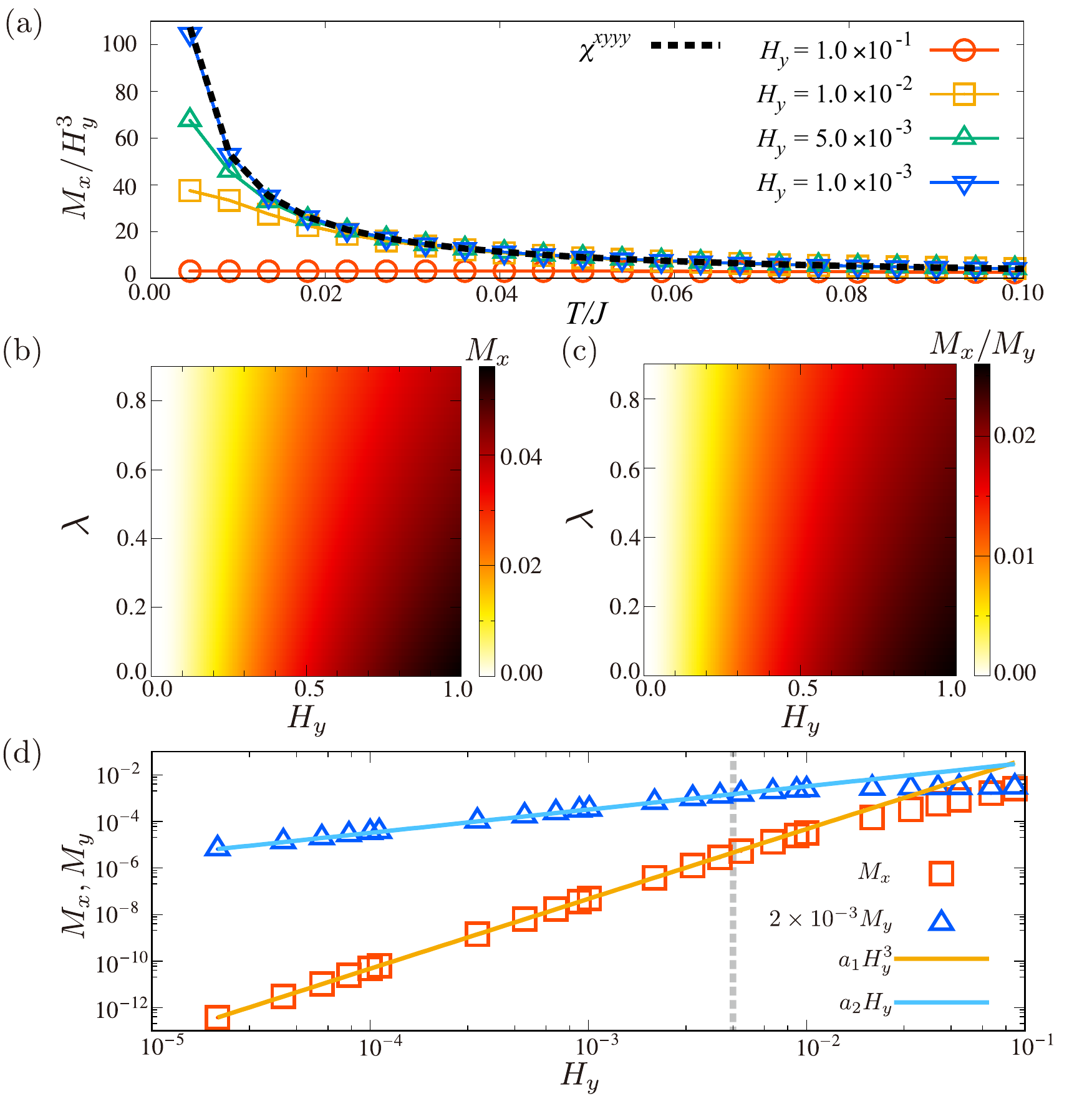}
\caption{(Color online) 
(a) $T$ dependence of $\mu_x / {H_y}^3$ for several $H_y$ at $\lambda= 0.2$~\cite{comment_data}. The dashed black curve represents
$\chi_{xyyy}$. 
Contour plots of (b) $M_x$ and (c) $M_x/M_y$ on the $H_y$--$\lambda$ plane at $T=0.1$.
(d) Magnetization process at $\lambda=0.2$ and $T=0.01$. 
The fitting lines by $a_1 H^3_y$ and $a_2 H_y$ with $a_1=47$ and $a_2=0.3231$ are also shown.
The vertical dashed line represents the magnetic field which deviates from $H^3_y$ ($H_y$) for $M_x$ $(M_y)$.}
\label{f4}
\end{figure}

So far, we have investigated the behavior of $\chi_{xyyy}$ by supposing the small magnetic field so that $\chi_{xyyy}$ is well scaled as $M_x/H_y^3$.
We here discuss the case for large $H_x$ at low temperatures beyond the perturbation regime. 
We show the $T$ dependence of $M_x/H^3_y$ at $\lambda = 0.2$ for several $H_y$ in Fig.~\ref{f4}(a). 
The data clearly indicate that $M_x$ for large $H_y$ deviates from the value expected from $\chi_{xyyy}$ (dashed black lines). 
A similar situation holds for different values of $\lambda$. 
We show the contour plots of $M_x$ and $M_x/M_y$ against $H_y$ and $\lambda$ at $T=0.1$ in Figs.~\ref{f4}(b) and \ref{f4}(c), respectively. 
One finds that the magnitude of the transverse magnetization $M_x$ reaches a few percent of that of longitudinal one $M_y$ for large $H_y$ because of the higher-order contributions. 
We also show the magnetization process at $\lambda=0.2$ and $T=0.01$, where $M_x$ ($M_y$)  deviates from $H^3_y$ ($H_y$) around $H_y \sim 0.004$.
Thus, a large magnetic field makes the detection of the transverse magnetization easier. 

In addition, one notices that the SOC dependence seems to be smaller.
Indeed, $\chi_{xyyy}$ remains nonzero even for $\lambda=0$. 
Such a situation is drastically changed when the different ferroaxial order parameter is considered.
For example, when we adopt the ordering of the spinless electric hexadecapole $Q^{\alpha}_{4z}$ instead of $G_z$, we find that $G_z=0$ and $\chi_{xyyy}$ becomes nonzero only for $\lambda \neq 0$ in spite of the same symmetry of $C_{\rm 4h}$.
This difference arises from the expressions of the order parameters: the ETD $G_z=(\bm{l}\times\bm{s})^z$ and the spinless electric hexadecapole $Q_{4z}^\alpha(\propto xy(x^2-y^2))$.
In other words, $G_z$ itself entangles spin and orbital components without the SOC, while $Q_{4z}^{\alpha}$ does not.
Thus, the SOC is essentially important for the $Q^{\alpha}_{4z}$ ordering. 

To summarize, we have investigated the nonlinear TMS under the ETD ordering by using the mean-field calculations and nonlinear response theory for the five $d$-orbital model on the square lattice. 
We have clarified the important model parameters and microscopic processes by analyzing the nonlinear Kubo formula.
We have found that the nonlinear TMS is caused by the collaborative effect between the ETD ordering and the tetragonal CEF.
Furthermore, we have shown that strong spin--orbital entanglement arising from the ETD order parameter plays an important role in enhancing the nonlinear transverse response. 
As the magnitude of the induced transverse magnetization is large enough to detect, it is expected to be observed in materials with the ETD (ferroaxial) moment.
The candidate materials are
$\text{Ca}\text{Mn}_7\text{O}_{12}$~\cite{ETD_CaMn7O12},
$\text{RbFe(MoO}_4)_2$~\cite{ETD_RbFeMoO42},
$\text{NiTiO}_3$~\cite{Hayashida_PhysRevMaterials.5.124409}, 
$\text{Ca}_5\text{Ir}_3\text{O}_{12}$~\cite{Hasegawa_doi:10.7566/JPSJ.89.054602, ETD_Ca5Ir3O12, hayami2023cluster}, and BaCoSiO$_4$~\cite{Xu_PhysRevB.105.184407}.

\acknowledgment
We are deeply grateful to A. Kirikoshi and R. Yambe for their valuable discussions.
This research was supported by JSPS KAKENHI Grants Numbers JP21H01037, JP22H04468, JP22H00101, JP22H01183, and by JST PRESTO (JPMJPR20L8). 

\bibliographystyle{JPSJ}
\bibliography{refETD}

\begin{thebibliography}{10}
\bibitem{comment_data}
The data denoted by the lines are calculated by $\Delta T=4.5 \times 10^{-3}$.
  The symbols are sparsely plotted for better visualization.
\end{thebibliography}
\clearpage
\end{document}